\documentclass{appolb}
\usepackage{graphicx}
\usepackage{braket}
\usepackage{dsfont}
\usepackage{amssymb,amsmath,amsfonts}
\usepackage{epstopdf}
\usepackage{xcolor}
\usepackage[utf8]{inputenc}
\usepackage{authblk}
\usepackage[nocompress]{cite}
\usepackage[square, numbers, comma, sort]{natbib}
\title{Shell-model calculation of isospin-symmetry breaking correction to superallowed Fermi beta-decay%
\thanks{The XXIII Nuclear Physics Workshop "Marie \& Pierre Curie", Kazimierz, Poland, September 27 to 2 October, 2016 }}%
\author[1]{L. Xayavong}
\author[1]{N. A. Smirnova}
\author[1,2]{M. Bender}
\author[2,3]{K. Bennaceur}

{\affil[1]{CENBG (CNRS/IN2P3 -- Universit\'e de Bordeaux), 33175 Gradignan cedex, France} 
\affil[2]{Univ Lyon, Université Lyon 1, CNRS/IN2P3, IPNL, F-69622 Villeurbanne, France}
\affil[3]{Department of Physics, PO Box 35 (YFL), FI-40014 University of Jyväskylä, Finland}

\begin{document}
 \eqsec  

\maketitle
\begin{abstract}

   We investigate  the radial-overlap part of the isospin-symmetry breaking correction to superallowed $0^+ \to 0^+$ $\beta $-decay
   using the shell-model approach similar to that of Refs.~\cite{ToHa1977,OrBr1989}. 
   The 8 $sd$-shell emitters with masses between $A$=22 and $A$=38 have been re-examined. 
   The Fermi matrix element is evaluated with realistic spherical single-particle wave functions, obtained from
   spherical Woods-Saxon (WS) or Hartree-Fock (HF) potentials, fine-tuned to reproduce the experimental data on charge radii and 
   separation energies for nuclei of interest. 
   The elaborated adjustment procedure removes any sensitivity of the correction to a specific parametrisation of the WS potential or 
   to various versions of the Skyrme interaction. 
   The present results are generally in good agreement with those reported in Refs.~\cite{ToHa2002,OrBr1985}. 
   At the same time, we find that the 
   calculations with HF wave functions result in systematically lower values of the correction.

\end{abstract}
  
\section{Physics of superallowed $\beta$ decay} 

It has been pointed out that the superallowed $0^+\to 0^+$ nuclear beta decay provides an excellent tool to probe the fundamental symmetries 
underlying the Standard Model of electroweak interaction, 
including the Conserved Vector Current (CVC) hypothesis and the unitarity of the Cabibbo-Kobayashi-Maskawa (CKM) quark-mixing matrix. 
According to the CVC hypothesis, the corrected $\mathcal{F}t$ value should relate to $G_V$, a fundamental
vector coupling constant for a semi-leptonic decay, and thus be constant for
all emitters. Traditionally, this relation is expressed
\begin{equation}\label{Ft}
\mathcal{F}t = ft(1+\delta'_R)(1+\delta_{NS}-\delta_C) = \frac{K}{2 \,G_V^2 \,(1+\Delta_R^V)}, 
\end{equation}
where 
$K/(\hbar c)^6 = 2\pi^3\ln(2)\hbar/(m_ec)^5 =$ $(8120.2716\pm0.012)\times 10^{-10}$~GeV$^{-4}$sec, 
$\Delta_R^V$, $\delta'_R$ and $\delta_{NS}$ are transition-independent, transition-dependent and nuclear-structure dependent parts
of a radiative correction~\cite{ToHa2002}, $\delta_C$ is an isospin-symmetry-breaking correction,
defined as a deviation of the Fermi matrix element squared from its model-independent value~:
\begin{equation}\label{c}
|M_F|^2=|M_F^0|^2(1-\delta_C), 
\end{equation}
with 
$|M_F^0|=\sqrt{ T(T+1)-T_{zi}T_{zf} }$ .  

The quantity $ft$ is determined experimentally by measuring the partial half-life, 
the $Q_{EC}$ value and the Fermi branching ratio.
The most recent survey of world data~\cite{HaTo2015} finds 14 of 
these superallowed transitions with measured $ft$ values known to 0.1\% precision or better. 
If the CVC holds, one can thus extract $G_V$.
By comparing it to the vector coupling constant from a muon decay, the CKM mixing matrix 
element between $u$ and $d$ quarks, $|V_{ud}|$ can be determined, providing a precise test of the
unitarity condition of the CKM matrix.

On the theoretical side, there is still no consensus between various calculations of $\delta_C$ (see Ref.~\cite{HaTo2015}
for a recent review). 
The present work explores the differences between shell-model calculations supplemented
by WS or HF radial wave functions in comparison with the previous studies~\cite{OrBr1985,ToHa2002}.

\section{Shell-model description of the isospin correction}

Within the shell model, the Fermi matrix element of the $\beta^+$ decay between an initial $|i \rangle $and final $|f \rangle $ many-body states
can be written as 
\begin{equation}\label{MF}
M_F = \braket{f |T_+| i} = \sum_\alpha\sum_\pi \braket{ f|a_{\alpha_n}^\dagger \ket{\pi}\bra{\pi} a_{\alpha_p}| i} \braket{\alpha_n| \hat t_+|\alpha_p }^\pi \, ,
\end{equation}
where $\alpha $ denotes a full set of spherical quantum numbers of a single-particle state, $\pi $ refers to a complete set of states
of an $(A-1)$ nucleus compatible with the angular momentum and parity conservation, and
 $\braket{\alpha_n |\hat t_+|\alpha_p }^\pi$ is the single-particle matrix element of the isospin operator~\footnote{We neglect the radial excitations 
as was pointed out by Miller and Schwenk~\cite{MiSch2008}} between proton and neutron radial wave functions:
\begin{equation}
\braket{\alpha_n |\hat t_+|\alpha_p }^\pi  = \Omega^\pi_\alpha = \int_0^\infty R^\pi_{\alpha_n}(r) R^\pi_{\alpha_p}(r) r^2 dr.
\end{equation}
For harmonic oscillator functions, the latter is equal to one. To overcome this artefact,
we have to replace the harmonic oscillator radial wave functions by realistic radial wave functions 
obtained from a spherically-symmetric WS or self-consistent HF potential.
A sum over intermediate states $\pi $ in Eq.~(\ref{MF}) allows us to go beyond the closure approximation and
take into account the dependence of $\Omega_\alpha^\pi$ on the excitation energies of the intermediate states, $E_{\pi }$.
For each $E_{\pi }$, we fine-tune our potential so that the individual energies of valence space orbitals match
experimental proton or neutron separation energies.

Substituting Eq.~\eqref{MF} into Eq.~\eqref{c}, we obtain a suitable expression for $\delta_{C}$, as a sum of two terms,
$\delta_C \approx  \delta_{RO} + \delta_{IM}$. 
The first term, $\delta_{RO}$, is the contribution due to the
deviation from unity of the overlap integral between the radial parts of the proton and neutron single-particle
wave functions. It is called a radial-overlap correction and can be expressed as
\begin{equation}\label{RO}
\delta_{RO} = \frac{2}{M_F^0} \sum_\alpha \sum_\pi \braket{ f||a_{\alpha_n}^\dagger ||\pi}^T  \braket{i||a^\dagger_{\alpha_p}|| \pi}^T (1-\Omega_\alpha^\pi),
\end{equation}
where the reduced matrix elements, $\braket{ f||a_{\alpha_n}^\dagger ||\pi}^T$ and $ \braket{i||a^\dagger_{\alpha_p}|| \pi}^T$ are related 
to the spectroscopic amplitudes~\cite{OrBr1985} for neutron and proton pick-up respectively. 
The superscript $T$ means that these quantities are computed with an isospin-invariant effective interaction. 

{The other term, $\delta_{IM}$, is the so-called isospin-mixing correction~\cite{OrBr1989,ToHa2008},
arising due to the isospin-mixing in many-body configurations of the initial and final states. 
It is obtained from the shell-model diagonalisation using a charge-dependent two-body effective interaction
and is expressed as}
\begin{equation}
\delta_{IM} = \frac{2}{M_F^0} \sum_\alpha [\braket{ f|a_{\alpha_n}^\dagger  a_{\alpha_p}| i}^T - \braket{ f|a_{\alpha_n}^\dagger  a_{\alpha_p}| i}] \, .
\end{equation}
In this work, we focus only on the radial overlap correction, calculating it within the shell model in
combination with the realistic radial wave functions
obtained from a WS or Skyrme-HF single-particle potential.

\section{Results for $\delta_{RO}$ and discussions}

The radial overlap correction, $\delta_{RO}$ has been evaluated using the procedures outlined in the previous section. 
For this study, we choose only $sd$-shell emitters 
which are well described by the so-called {\it universal} $sd$ interactions --- USD, USDA/B~\cite{USD,USDab}.
They include $^{22}\mbox{Mg}$, $^{26}\mbox{Al}$, 
$^{26}\mbox{Si}$, $^{30}\mbox{S}$, $^{34}\mbox{Cl}$, $^{34}\mbox{Ar}$, $^{38}\mbox{K}$ and $^{38}\mbox{Ca}$. 
Six of these transitions are used to deduce the most precise $\mathcal{F}t$ value, while 
the decays of $^{26}\mbox{Si}$ and $^{30}\mbox{S}$ are expected
to be measured with an improved precision in future radioactive-beam facilities. 

The shell-model calculations have been performed in the full $sd$ shell, 
using NuShellX@MSU code~\cite{nushell}. 
To get convergence, up to 100 intermediate states of each spin have been taken into account in Eq.~\eqref{RO}.

\begin{figure*}[ht!]
\centering
\includegraphics[scale=1.4]{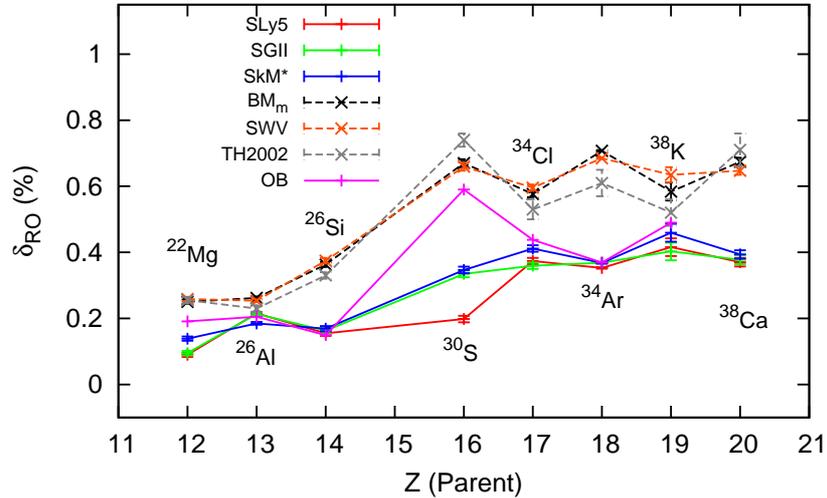}
\caption{(Color online). Calculated values for the radial overlap correction, $\delta_{RO}$ (\%) in comparison with the
results of Ormand and Brown (OB)~\cite{OrBr1985,OrBr1989} and those of Towner and Hardy (TH2002)~\cite{ToHa2002}. }
		\label{f1}
\end{figure*}

Figure~\ref{f1} shows the results for $\delta_{RO}$ obtained with either WS or HF single-particle wave functions. 
The WS results have been computed using two different parametrisations.
One of them is that of Schwierz, Wiedenhöver and Volya (SWV)~\cite{SWV}, while  
the other is that of Bohr and Mottelson~\cite{BohrMott}, modified as proposed in Ref.~\cite{TALENT} 
and denoted as $BM_m$. 
Important to note that the Coulomb and the charge-symmetric
isovector terms are the only sources of the difference
between proton and neutron single-particle wave functions. 
In the present study, we assume that the
charge-symmetry breaking and all other deficiencies
of the WS potential can be cured by readjustment of the well depth case-by-case 
to reproduce experimental proton and neutron separation energies.
The length parameter of the central term  was determined
from a condition that the charge density constructed from
the proton radial wave functions yields a root-mean-charge radius in agreement with the experimental value
measured by electron scattering~\cite{radius} or by isotope-shift estimation~\cite{ToHa2002}. 

The spherical HF calculations have been performed with three different Skyrme forces, 
namely,  SGII~\cite{SGII} and SkM*~\cite{SKM*} and SLy5~\cite{SLY5}.
While SGII and SkM* were already used in Ref.~\cite{OrBr1985}, SLy5 is a more recent parametrisation by the Saclay-Lyon collaboration. 
It was constructed to reproduce various bulk nuclear properties and selected properties of a number of doubly magic nuclei, without $^{16}$O. 
Since the nuclei of interest are open-shell systems, we have assumed a uniform occupation 
of a last occupied, partly filled orbital.
We have checked that shell-model occupation numbers for initial and final $0^+$ states, 
obtained from the diagonalisation, produce very similar results. 
The central part of the self-consistent potential of the parent and daughter nuclei was scaled in order to reproduce 
experimental proton and neutron separation energies, respectively. 
We have tested that this scaling only little influences the charge radii of nuclei considered, 
which stay in very good agreement with experiment. 
The Coulomb exchange term was accounted within the Slater approximation.
Our preliminary results~\cite{LaPhD} show that its exact calculation only marginally affects the $\delta_{RO}$ value.

As is seen from the figure, all our WS results are quite close to each other, 
indicating that the correction $\delta_{RO}$ is not very sensitive 
to a particular choice of the WS potential parameters. 
In general, they are in fair agreement with the shell-model plus WS calculation of Towner and Hardy in 2002~\cite{ToHa2002}, 
except for $^{34}\mbox{Ar}$ and 
$^{38}\mbox{K}$ because we used new experimental data for the charge radius~\cite{radius}. 
We do not compare our present results with the latest calculation of Towner and Hardy~\cite{ToHa2008},
performed with the inclusion of the orbitals outside the valence space. The work in this direction is in progress.

For the HF case, we find that the correction only little depends on a particular version of the Skyrme force,
except for $^{30}$S.
Overall, our results are consistent with those of Ormand and Brown~\cite{OrBr1989}, again with the exception 
of $^{30}\mbox{S}$. 
In the case of $^{30}$S, the correction $\delta_{RO}$ is dominated by 
the $2s_{1/2}$ state, in which the centrifugal barrier is not present, and thus the radial wave function is very sensitive 
to the fine details of the mean field. We note that SLy5 interaction results in a considerably smaller $\delta_{RO}$ value
compared to SGII and SkM*.

We do not confront our results to the most recent calculation of $\delta_{RO}$ with Skyrme-HF wave 
functions carried out by Hardy and Towner in 2009~\cite{HaTo2009}. 
Unlike the standard HF procedure, they performed 
a single calculation for the nucleus with $(A-1)$ nucleons and $(Z-1)$ protons, and 
then used the proton and the neutron eigenfunctions from the same calculation to compute radial integrals. 
{Since Koopman's theorem is not fully respected by such HF calculations, 
in particular, with a density-dependent effective interaction, we do not consider their protocol to be well justifiable.} 

The $\delta_{RO}$ values obtained with HF wave functions are seen to be systematically smaller
than those obtained with WS wave functions. The reason can be easily understood.
The Skyrme interaction is usually supposed to be isospin invariant. 
However, the presence of the Coulomb term causes a difference between proton and neutron densities, 
inducing an isovector term in the self-consistent mean-field potential~\cite{DG,OrBr1985}. 
That term tends to counter Coulomb repulsion, therefore reducing $\delta_{RO}$.  
It will be interesting to study whether charge-symmetry (CSB) and charge-independence breaking (CIB) 
terms in a conventional isospin-invariant Skyrme interaction affect the value of the correction.

\section*{Acknowledgments}

We are grateful to B. Blank for stimulating discussions.
L.~Xayavong thanks Université de Bordeaux for a Ph.D. fellowship. 
The work was supported by the CFT (IN2P3/CNRS, France), AP théorie 2014--2016.


\end{document}